\documentclass[aps,superscriptaddress,12pt,dvipdfm]{revtex4}                 
\usepackage{graphics,graphicx,dcolumn,bm,fleqn,epic,eepic,float}
\usepackage{amssymb,amsmath,multirow,rotate,color,float}
\usepackage{times}
\usepackage{verbatim}
\definecolor{red}{rgb}{1,0,0}

\definecolor{blue}{rgb}{0,0,1}

\begin{document}
%%%%%%%%%%%%%%%%%%%%%%%%%%%%%%%%%%%%
%\draft
%\preprint{HEP/123-qed}
%%%HEADINGS
\title{Understanding individual human mobility patterns}

\date{\today}
\author{Marta C.~Gonz\'alez}
%\email[Email: ]{marta@ica1.uni-stuttgart.de}
%\homepage[URL: ]{http://www.ica1.uni-stuttgart.de/~marta}
\affiliation{Center for Complex Network Research and Department of
Physics and Computer Science, University of Notre Dame, Notre Dame IN 46556.}
\affiliation{Center for Complex Network Research and Department of Physics, Biology and
Computer Science, Northeastern University, Boston MA 02115.}
\author{C\'{e}sar A.~Hidalgo}
\affiliation{Center for Complex Network Research and Department of
Physics and Computer Science, University of Notre Dame, Notre Dame IN 46556.}
\author{Albert-L\'{a}szl\'{o} Barab\'{a}si}
%\homepage[URL: ]{http://www.ica1.uni-stuttgart.de/~hans}
\affiliation{Center for Complex Network Research and Department of
Physics and Computer Science, University of Notre Dame, Notre Dame IN 46556.}
\affiliation{Center for Complex Network Research and Department of Physics,
  Biology and Computer Science, Northeastern University, Boston MA 02115.}
\affiliation{Center for Cancer Systems Biology, Dana Farber Cancer Institute,
  Boston, MA 02115.}

%%%%%%%%%%%%%%%%%%%%%%%%%%%%%%%%%%%%%%%
%\begin{abstract}
%\end{abstract}
%%%%PACS e Keywords
%\pacs{
%89.65.Ef
%02.50.Le
%64.60.Ak
%89.75.Hc
%89.65.Ef Social organization
%02.50.Le Game Theory
%64.60.Ak Percolation
%89.75.Hc Networks
%}

%%%%%%%%%%%%%%%%%%%%%%%%%%%%%%%%%%%%%%%%%%%%%%%%%%%%%%%%%%%%%%
%%%%%%%%%%%%%%%%%%%%%%%% TEXT %%%%%%%%%%%%%%%%%%%%%%%%%%%%%%%%
%%%%%%%%%%%%%%%%%%%%%%%%%%%%%%%%%%%%%%%%%%%%%%%%%%%%%%%%%%%%%%
%64.60.Ak Percolation
%89.75.Hc Networks
%%%%%%%%%%%%%%%%%%
\maketitle
%%%%%%%%%%%%%%%%%%%%%%%%%%%%%%%%%%%%%%%%%%%%%%%%%%%%%%%%%%%%%%
%%%%%%%%%%%%%%%%%%%%%%%% TEXT %%%%%%%%%%%%%%%%%%%%%%%%%%%%%%%%
%%%%%%%%%%%%%%%%%%%%%%%%%%%%%%%%%%%%%%%%%%%%%%%%%%%%%%%%%%%%%%
{\bf
Despite their importance for urban planning~\cite{Horner}, traffic
forecasting~\cite{Kitamura}, and the spread of biological~\cite{Vespignani,Zoltan,Hufnagel} and
mobile viruses~\cite{Kleinberg}, our understanding of the
basic laws governing human motion remains limited thanks to the lack of tools
to monitor the time resolved location of individuals.
Here we study the trajectory of $100,000$ anonymized mobile
phone users whose position is tracked for a six month period.
We find that in contrast with the random trajectories predicted
by the prevailing L\'{e}vy flight and random walk models~\cite{Geisel}, human trajectories show a high degree of temporal and spatial
regularity, each individual being characterized by a time independent
characteristic length scale and a significant probability to return to a few
highly frequented locations. After correcting
for differences in travel distances and the inherent anisotropy of each
trajectory, the individual travel patterns collapse into a single spatial
probability distribution, indicating that despite the diversity of their
travel history, humans follow simple reproducible patterns.
This inherent similarity in travel patterns could impact all phenomena
driven by human mobility, from epidemic prevention to
emergency response, urban planning and agent based modeling.}

Given the many unknown factors that influence a population's
mobility patterns, ranging from means of transportation
to job and family imposed restrictions and priorities, human trajectories are often
approximated with various random walk or diffusion models~\cite{Havlin,Geisel}.
Indeed, early measurements on albatrosses, bumblebees, deer and monkeys~\cite{Stanley,Miramontes} and more recent ones on marine predators~\cite{Sims}
suggested that animal trajectory is approximated by a
L\'evy flight~\cite{Klafter,Mantegna},
a random walk whose step size $\Delta r$ follows a
power-law distribution $P(\Delta r) \sim \Delta r^{-(1+\beta)}$ with $\beta<2$.
While the L\'{e}vy statistics for some animals require further study~\cite{Stanley_new},
Brockmann {\it et al.}~\cite{Geisel} generalized this finding to humans,
documenting that the distribution of distances
between consecutive sightings of nearly half-million bank notes is fat tailed.
Given that money is carried by individuals, bank note dispersal is a proxy for
human movement, suggesting that human trajectories are best modeled
as a continuous time random walk with fat tailed displacements and
waiting time distributions~\cite{Geisel}. A particle
following a L\'{e}vy flight has a significant probability to travel very long
distances in a single step~\cite{Klafter,Mantegna}, which appears to be
consistent with human travel patterns: most of the time we travel
only over short distances, between home and work, while occasionally we take longer trips.

Each consecutive sightings of a bank note reflects
the composite motion of two or more individuals, who owned the
bill between two reported sightings. Thus it is not clear if the observed
distribution reflects the motion of individual users, or some hitero unknown
convolution between population based heterogeneities and individual human
trajectories. Contrary to bank notes, mobile phones
are carried by the same individual during his/her
daily routine, offering the best proxy
to capture individual human trajectories~\cite{Sohn,Onnela,Gonzalez,Palla,Hidalgo}.

\begin{figure}[htb]
\begin{center}
\includegraphics[width=12.0cm]{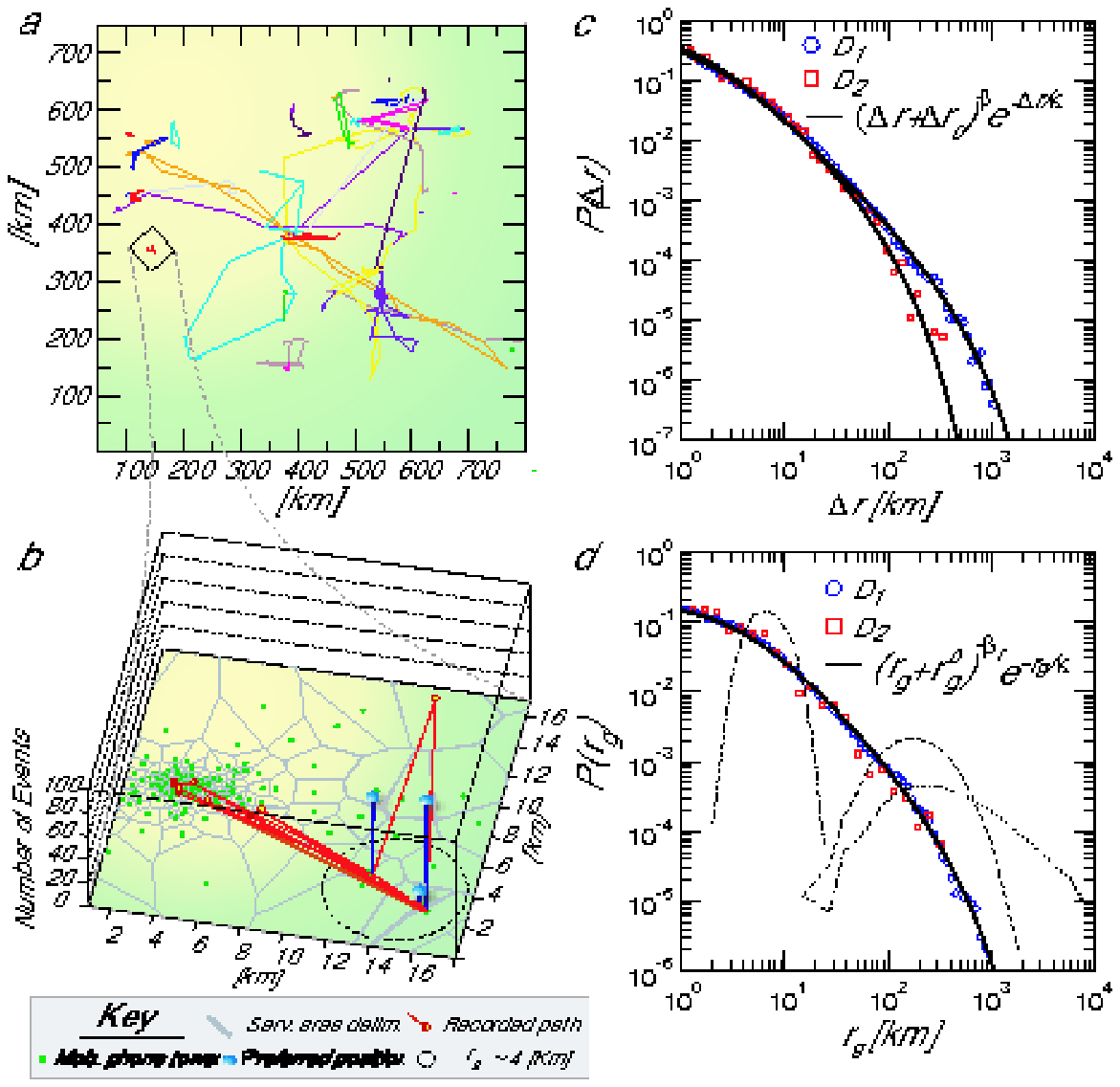}
\end{center}
\caption{
\baselineskip=12pt
{\bf Basic human mobility patterns}.
{\bf a,} Week-long trajectory of $40$ mobile phone users indicate
that most individuals travel only over short distances, but a few regularly
move over hundreds of kilometers. Panel {\bf b,} displays the detailed trajectory of a single user. The
different phone towers are shown as green dots, and the Voronoi lattice in grey
marks the approximate reception area of each tower. The dataset studied by
us records only the identity of the closest tower to a mobile user, thus
we can not identify the position of a user within a Voronoi cell.
The trajectory of the user shown in {\bf b} is constructed from
$186$ two hourly reports, during which the user visited
a total of $12$ different locations (tower vicinities). Among these,
the user is found $96$ and $67$ occasions in the two most preferred
locations, the frequency of visits for each location being shown as a vertical
bar.
The circle represents the radius of gyration centered
in the trajectory's center of mass.
{\bf c,} Probability density function $P(\Delta r)$ of travel distances obtained
for the two studied datasets $D_{1}$ and $D_{2}$. The solid line indicates a
truncated power law whose parameters are provided in the text (see Eq.~\ref{eq:trunc}).
{\bf d,} The distribution  $P(r_{g})$ of the radius of gyration
measured for the users, where $r_{g}(T)$ was measured after $T=6$ months
of observation. The solid line represent a similar
truncated power law fit (see Eq.~\ref{eq:prg}).
The dotted, dashed and dot-dashed curves show $P(r_{g})$  obtained from
the standard null models ($RW$, $LF$ and $TLF$), where for the $TLF$
we used the same step size distribution as the one
measured for the mobile phone users.}
\label{fig1}
\end{figure}

 We used two data sets to explore the mobility pattern of individuals.
The first ($D_{1}$) consists of the mobility patterns recorded over
a six month period for $100,000$ individuals selected randomly
from a sample of over 6 million anonymized mobile phone users.
Each time a user initiates or receives a call or SMS,
the location of the tower routing the communication is
recorded, allowing us to reconstruct the user's time resolved trajectory
(Figs.~\ref{fig1}a and b).
The time between consecutive calls follows a bursty
pattern~\cite{BarabasiNature} (see Fig. S1 in the SM),
indicating that while most consecutive calls are placed soon after a previous call,
occasionally there are long periods without any call
activity. To make sure that the obtained results are not affected by the irregular
call pattern, we also study a data set ($D_{2}$) that
captures the location of $206$ mobile phone users, recorded every two hours
for an entire week.
In both datasets the spatial resolution is determined by the local density of the more than
$10^4$ mobile towers, registering movement only when the user moves between
areas serviced by different towers. The average service area of each tower is approximately
$3\:\textrm{km}^{2}$ and over $30\%$ of the towers cover an area of $1\:\textrm{km}^{2}$ or less.

To explore the statistical properties of the population's mobility patterns
we measured the distance between user's positions at
consecutive calls, capturing $16,264,308$ displacements
for the $D_1$ and $10,407$ displacements for the $D_2$
datasets. We find that the distribution of displacements over all users
is well approximated by a truncated power-law
\begin{equation}
P(\Delta r) = (\Delta r+\Delta r_{0})^{-\beta}\exp{(-\Delta r/\kappa)},
\label{eq:trunc}
\end{equation}
with $ \beta = 1.75 \pm 0.15$, $\Delta r_{0} = 1.5$ km and
cutoff values $\kappa|_{D_{1}} = 400$ km, and $\kappa|_{D_{2}} = 80$ km
(Fig.~\ref{fig1}c, see the SM for statistical validation).
Note that the observed scaling exponent is not far
from $\beta_{B}=1.59$ observed in Ref.~\cite{Geisel} for
bank note dispersal, suggesting that the two distributions
may capture the same fundamental mechanism
driving human mobility patterns. 

Equation (\ref{eq:trunc}) suggests that human motion follows
a truncated L\'{e}vy flight~\cite{Geisel}.
Yet, the observed shape of $P(\Delta r)$
could be explained by three distinct
hypotheses: A. Each individual
follows a L\'{e}vy trajectory with jump size distribution
given by (\ref{eq:trunc}). B. The observed distribution captures a population based
heterogeneity, corresponding to the inherent differences between individuals.
C. A population based heterogeneity coexists with
individual L\'{e}vy trajectories, hence (\ref{eq:trunc}) represents a
convolution of hypothesis $A$ and $B$.

To distinguish between hypotheses A, B and C we calculated the
radius of gyration for each user (see Methods), interpreted as the
typical distance traveled by user $a$ when observed up to time $t$ (Fig.~\ref{fig1}b). Next, we determined the radius of gyration distribution
$P(r_{g})$ by calculating $r_{g}$ for all users
in samples $D_{1}$ and $D_{2}$, finding that they also can be approximated
with a truncated power-law
\begin{equation}
P(r_{g}) = (r_{g}+r_{g}^{0})^{-\beta_{r}}\exp{(-r_{g}/\kappa)},
\label{eq:prg}
\end{equation}
with $r_{g}^{0} = 5.8$ km, $\beta_{r}=1.65 \pm 0.15$  and $\kappa = 350$ km
(Fig.~\ref{fig1}d, see SM for statistical validation).
L\'{e}vy flights are characterized by a high degree of intrinsic
heterogeneity, raising the possibility that (\ref{eq:prg})
could emerge from an ensemble of identical agents,
each following a L\'{e}vy trajectory. Therefore, we determined $P(r_{g})$ for an ensemble
of agents following a Random Walk ($RW$), L\'evy-Flight ($LF$)
or Truncated L\'evy-Flight ($TLF$) (Figure~\ref{fig1}d)~\cite{Klafter,Mantegna,Havlin}.
We find that an ensemble of L\'{e}vy agents display
a significant degree of heterogeneity in $r_{g}$, yet is not
sufficient to explain the truncated power law distribution $P(r_{g})$
exhibited by the mobile phone users.
Taken together, Figs.~\ref{fig1}c and d suggest that the difference in the
range of typical mobility patterns of individuals ($r_{g}$) has a strong
impact on the truncated L\'{e}vy behavior seen in (\ref{eq:trunc}), ruling out hypothesis A.

\begin{figure*}[htb]
\begin{center}
\includegraphics*[width=8.9cm]{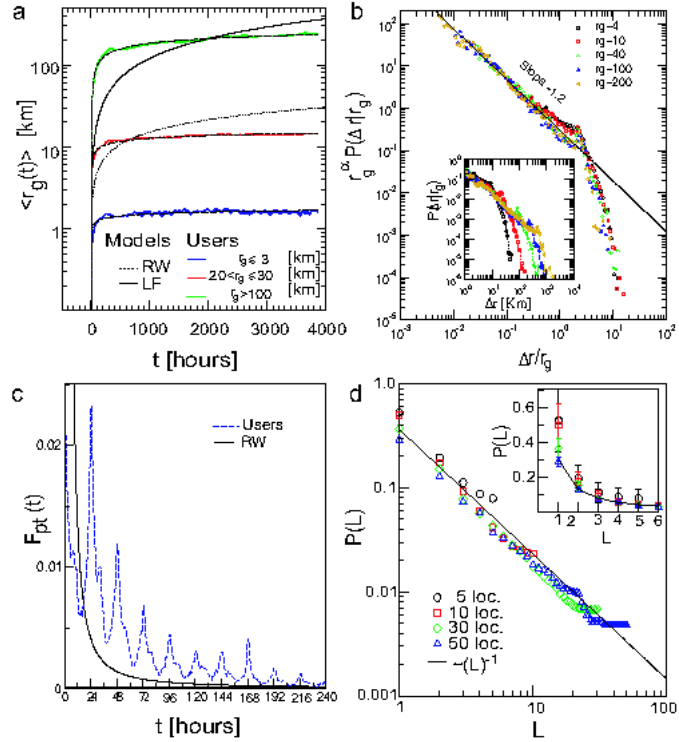}
\end{center}
\caption{\baselineskip=12pt
{\bf The bounded nature of human trajectories.}
{\bf a,} Radius of gyration, $\langle r_{g}(t) \rangle$ vs time for mobile
phone users separated in three groups according to their final $r_{g}(T)$,
where $T=6$ months.
The black curves correspond to the analytical predictions for the random
walk models, increasing in time as $\langle r_{g}(t) \rangle|_{LF,TLF}
\sim t^{3/2+\beta}$ (solid), and $\langle r_{g}(t) \rangle|_{RW} \sim
t^{0.5}$ (dotted). The dashed curves corresponding to a logarithmic
fit of the form $A + B \ln(t)$, where $A$ and $B$ depend on $r_{g}$.
{\bf b,} Probability density function of
individual travel distances $P(\Delta r|r_{g})$ for users with $r_{g} = 4$,
$10$, $40$, $100$ and $200$ km. As the inset shows, each
group displays a quite different $P(\Delta r|r_{g})$ distribution. After
rescaling the distance and the distribution with $r_{g}$ (main panel),
the different curves collapse. The solid line (power law) is
shown as a guide to the eye.
{\bf c,} Return probability distribution, $F_{pt}(t)$. The prominent
peaks capture the tendency of humans to regularly return to the
locations they visited before, in contrast with the smooth asymptotic behavior
$\sim 1/(t \ln(t)^{2})$ (solid line) predicted for random walks. {\bf d,} A
Zipf plot showing the frequency of visiting different locations.
The symbols correspond to users that have been observed
to visit $n_{L}=5$, $10$, $30$, and $50$ different locations.
Denoting with ($L$) the rank of the location listed in the
order of the visit frequency, the data is well approximated by $R(L) \sim L^{-1}$.
The inset is the same plot in linear scale, illustrating that $40\%$ of the
time individuals are found at their first two preferred locations.}
\label{fig2}
\end{figure*}

If individual trajectories are described by a $LF$ or $TLF$, then
the radius of gyration should increase in time
as $r_{g} (t) \sim t^{3/(2+\beta)}$~\cite{Book,Redner} while
for a $RW$ $ r_{g}(t) \sim t^{1/2}$.
That is, the longer we observe a
user, the higher the chances that she/he will travel to areas
not visited before.
To check the validity of these predictions we measured
the time dependence of the radius of gyration for users whose gyration radius would be considered small ($r_{g}(T) \le 3$ km), medium
($20 < r_{g}(T) \le 30$ km) or large ($r_{g}(T)>100$ km) at the end of
our observation period ($T=6$ months). The results indicate that the time
dependence of the average radius of gyration of mobile phone users is better
approximated by a logarithmic increase, not only
a manifestly slower dependence than the one predicted by a power law, but one that may
appear similar to a saturation process (Fig.~\ref{fig2}a and Fig. S4).

In Fig.~\ref{fig2}b, we have chosen users with similar
asymptotic $r_{g}(T)$ after $T=6$ months, and measured the jump size
distribution $P(\Delta r|r_{g})$ for each group. As the inset of Fig.~\ref{fig2}b shows,
users with small $r_{g}$ travel mostly over small distances, whereas
those with large $r_{g}$ tend to display a combination of many small and a
few  larger jump sizes.
Once we rescale the distributions with $r_{g}$ (Fig.~\ref{fig2}b),
we find that the data collapses into a single curve, suggesting that
a single jump size distribution characterizes all users, independent of their $r_{g}$.
This indicates that $P(\Delta r|r_{g}) \sim r_{g}^{-\alpha} F(\Delta r/r_{g})$, where
$\alpha \approx 1.2 \pm 0.1$ and $F(x)$ is an $r_{g}$ independent function with
asymptotic behavior $F(x < 1) ~\sim x^{-\alpha}$ and rapidly decreasing for $x \gg 1$.
Therefore the travel patterns of
individual users may be approximated by
a L\'{e}vy flight up to a distance characterized by $r_{g}$.
Most important, however, is the fact that the individual trajectories
are bounded beyond $r_{g}$, thus large displacements which are the
source of the distinct and anomalous nature of L\'{e}vy 
flights, are statistically absent. To understand the relationship between the
different exponents, we note that the measured probability
distributions are related by $P(\Delta r) = \int_{0}^{\infty} P(\Delta
r|r_{g})P(r_{g})dr_{g}$, which suggests (see SM) that up to the leading order
we have $\beta=\beta_{r}+\alpha-1$, consistent,
within error bars, with the measured exponents.
This indicates that the observed jump size distribution $P(\Delta {r})$ is in
fact the convolution between the statistics of individual trajectories $P(\Delta
r_{g}|r_{g})$ and the population heterogeneity $P(r_{g})$, consistent with
hypothesis C.

To uncover the mechanism stabilizing $r_{g}$ we measured the return probability for each individual
$F_{pt}(t)$ ~\cite{Redner}, defined as the probability
that a user returns to the position where it was first observed after $t$ hours
(Fig.~\ref{fig2}c). For a two dimensional random walk $F_{pt}(t)$ should
follow $\sim 1/(t \ln(t)^{2})$~\cite{Redner}.
In contrast, we find that the return probability is
characterized by several peaks at 24 h, 48 h, and 72 h,
capturing a strong tendency of humans to return to
locations they visited before, describing the recurrence and 
temporal periodicity inherent to human mobility~\cite{Schlich,Pentland}.

To explore if individuals return to the same location over
and over, we ranked each location based on the number of times
an individual was recorded in its vicinity, such that
a location with $L=3$ represents the
third most visited location for the selected individual.
We find that the probability of finding
a user at a location with a given rank $L$ is well approximated by $P(L)\sim 1/L$,
independent of the number of locations visited by the user
(Fig.~\ref{fig2}d).
Therefore people devote most of their time
to a few locations, while spending their
remaining time in $5$ to $50$ places, visited with diminished regularity.
Therefore, the observed logarithmic saturation of $r_{g}(t)$ is rooted in the high degree of regularity
in their daily travel patterns, captured by the high
return probabilities (Fig.~\ref{fig2}b) to a few highly frequented
locations (Fig.~\ref{fig2}d). 

\begin{figure}[htb]
\begin{center}
\includegraphics[width=16.0cm]{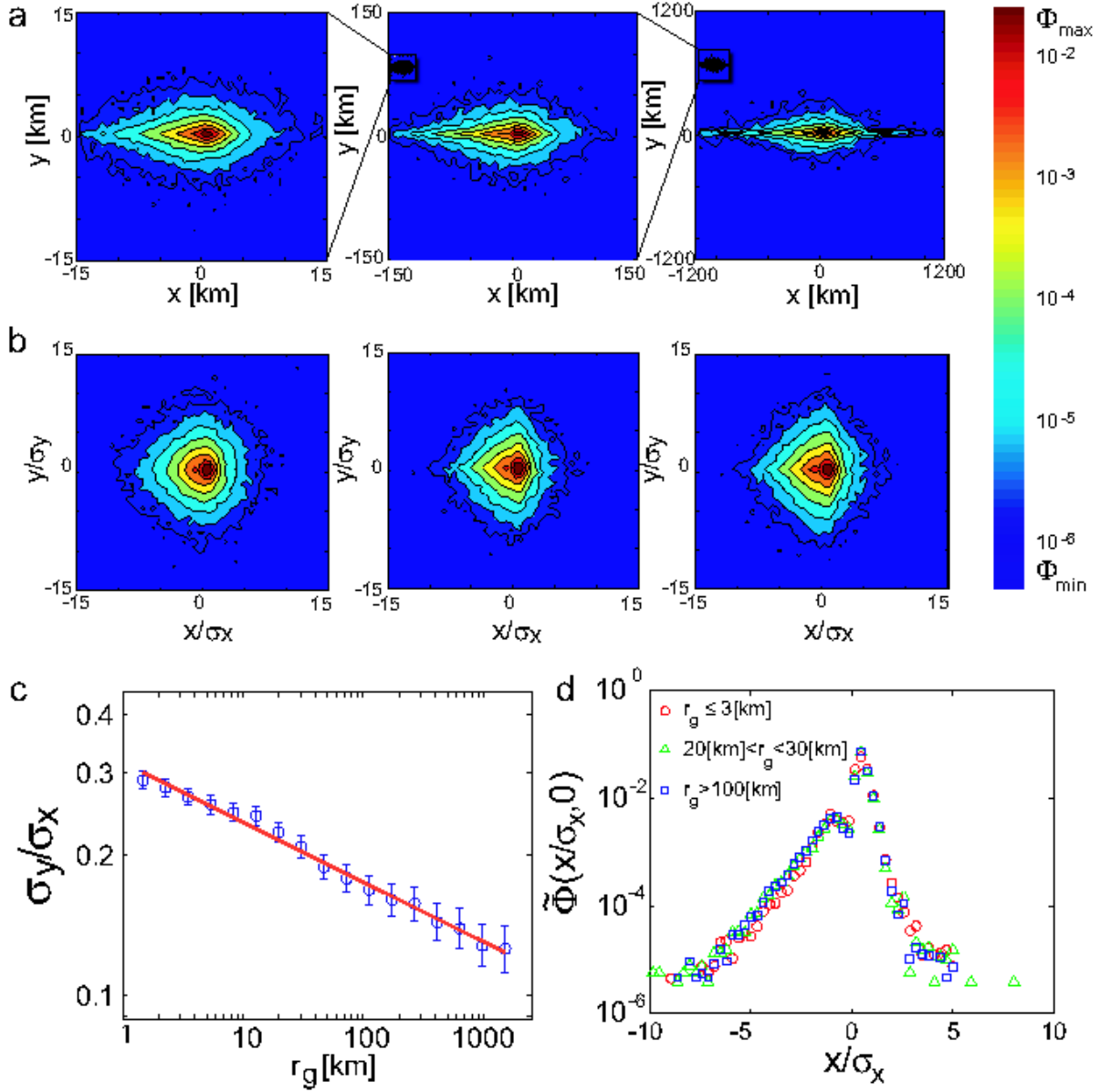}
\end{center}
\caption{\baselineskip=12pt
{\bf The shape of human trajectories.}
{\bf a,} The probability density
function $\Phi(x,y)$ of finding a mobile phone user
in a location $(x,y)$ in the user's intrinsic reference frame (see SM for details).
The three plots, from left to right, were generated for $10,000$ users with: $r_{g} \le 3$, $20 < r_{g} \le 30$
and $r_{g} > 100$ km. The trajectories become more anisotropic
as $r_{g}$ increases. {\bf b,} After scaling
each position with $\sigma_{x}$ and $\sigma_{y}$
the resulting $\tilde{\Phi}(x/\sigma_{x},y/\sigma_{y})$ has approximately
the same shape for each group. {\bf c,} The change in the shape of $\Phi(x,y)$
can be quantified calculating the isotropy ratio
$S \equiv \sigma_{y}/\sigma_{x}$ as a function of $r_{g}$,
which decreases as $S \sim r_{g}^{-0.12}$ (solid line). Error
bars represent the standard error.
{\bf d,} $\tilde{\Phi}(x/\sigma_{x},0)$ representing the
x-axis cross section of the rescaled distribution
$\tilde{\Phi}(x/\sigma_{x},y/\sigma_{y})$
shown in b.}
\label{fig3}
\end{figure}

An important quantity for modeling human mobility patterns is the probability
$\Phi_{a}(x,y)$ to find an individual $a$ in a given position ($x,y$).
As it is evident from Fig.~\ref{fig1}b,
individuals live and travel in different regions,
yet each user can be assigned to a well defined area,
defined by home and workplace,
where she or he can be found most of the time. We can compare the
trajectories of different users by diagonalizing each trajectory's
inertia tensor, providing
the probability of finding a user in a given position (see Fig.~\ref{fig3}a)
in the user's intrinsic reference frame (see SM for the details).
A striking feature of $\Phi(x,y)$ is its
prominent spatial anisotropy in this intrinsic reference frame
(note the different scales in Fig~\ref{fig3}a),
and we find that the larger an individual's $r_g$ the more pronounced is this anisotropy.
To quantify this effect we defined the
anisotropy ratio $S \equiv \sigma_y / \sigma_x$,
where $\sigma_{x}$ and $\sigma_{y}$ represent the standard deviation of
the trajectory measured in the user's intrinsic reference frame (see SM).
We find that $S$ decreases monotonically with
$r_{g}$ (Fig.~\ref{fig3}c), being well approximated with $S \sim r_{g}^{-\eta}$,
for $\eta \approx 0.12$. Given the small value of the scaling exponent,
other functional forms may offer an equally good fit, thus mechanistic models
are required to identify if this represents a
true scaling law, or only a reasonable approximation to the data.

To compare the trajectories of different users we remove the individual
anisotropies, rescaling each user trajectory with its respective $\sigma_x$ and $\sigma_y$. The rescaled
$\tilde{\Phi}(x/\sigma_{x},y/\sigma_{y})$ distribution
(Fig.~\ref{fig3}b) is similar for groups of users with considerably different
$r_{g}$, {\it i.e.}, after the anisotropy and the $r_{g}$ dependence is
removed all individuals appear to follow the same
universal $\tilde{\Phi}(\tilde{x},\tilde{y})$ probability distribution. This is particularly evident
in Fig.~\ref{fig3}d, where we show the cross section of
$\tilde{\Phi}(x/\sigma_{x},0)$
for the three groups of users, finding that apart from the noise in the
data the curves are indistinguishable.

Taken together, our results suggest that the L\'{e}vy
statistics observed in bank note measurements capture a convolution of the population
heterogeneity (\ref{eq:prg}) and the motion of individual users. 
Individuals display significant regularity, as they return to
a few highly frequented locations, like home or work. This
regularity does not apply to the bank notes: a bill always follows
the trajectory of its current owner, {\it i.e.} dollar bills diffuse, but humans do not.

The fact that individual trajectories are characterized 
by the same $r_{g}$-independent two dimensional probability distribution
$\tilde{\Phi}(x/\sigma_x, y/\sigma_y)$ suggests that key statistical
characteristics of individual trajectories are largely indistinguishable
after rescaling. 
Therefore, our results establish the basic ingredients of
realistic agent based models, requiring us to place users in number proportional with the population
density of a given region and assign each user an $r_g$ taken from the observed  $P(r_{g})$
distribution. Using the predicted anisotropic rescaling,
combined with the density function $\tilde{\Phi}(x,y)$,
whose shape is provided as Table 1 in the SM,
we can obtain the likelihood of finding a user
in any location. 
Given the known correlations between spatial proximity and social
links, our results could help quantify the role of space in network
development and evolution~\cite{Yook,Caldarelli,Mendes,Song,Gonzalez2} and
improve our understanding of diffusion processes~\cite{Havlin,Cecconi}.

\vspace{0.75in}

%\textbf{Acknowledgments}

We thank D. Brockmann, T. Geisel, J. Park, S. Redner, Z. Toroczkai and P. Wang
for discussions and comments on the manuscript.
This work was supported by the James S. McDonnell Foundation 21st Century Initiative in Studying
Complex Systems, the National Science Foundation within the DDDAS (CNS-0540348), ITR (DMR-0426737)
and IIS-0513650 programs, and the U.S. Office of Naval Research Award N00014-07-C.
Data analysis was performed on the Notre Dame Biocomplexity Cluster
supported in part by NSF MRI Grant No. DBI-0420980. C.A. Hidalgo acknowledges
support from the Kellogg Institute at Notre Dame.

\vspace{0.5in}
{\bf Supplementary Information} is linked to the online version  of
the paper at www.nature.com/nature. 

\vspace{0.5in}

{\bf Author Information} Correspondence and requests
for materials should be addressed to A.-L.B.
(e-mail: alb@nd.edu)
\newpage

%%%%%%%%%%%%%%%%%%%%%%%%%%%%%%%%%%%%%%%%%%%%%%%%%%%%%%%%%%%%%%%%%%%%%%%%%%%
%%%%%%%%%%%%%%REFERENCIAS %%%%%%%%%%%%%%%%%%%%%%%%%%%%%%%%%%%%%%%%%%%%%%%%%
%%%%%%%%%%%%%%%%%%%%%%%%%%%%%%%%%%%%%%%%%%%%%%%%%%%%%%%%%%%%%%%%%%%%%%%%%%%

%%%%%%%%%%%%%%%%%%%%%%%%%%%%%%%%%%%%%%%%%%%%%%%%%%%%%%%%%%%%%%%%%%%%%%%%%%%
\newpage
%%%%%%%%%%%%%%%%%%%%%%%%%%%%%%%%%%%%%%%%%%%%%%%%%%%%%%%%%%%%%%%%%%%%%%%
%\begin{comment}
%\end{comment}
%%%%%%%%%%%%%%%%%%%%%%%%%%%%%%%%%%%%%%%%%%%%%%%%%%%%%%%%%%%%%%%%%%%%%%%
%%%%%%%%%%%%%%%%%%%%%%%%%%%%%%%%%%%%%%%%%%%%%%%%%%%%%%%%%%%%%%%%%%%%%%%
%\begin{comment}
%\end{comment}
%%%%%%%%%%%%%%%%%%%%%%%%%%%%%%%%%%%%%%%%%%%%%%%%%%%%%%%%%%%%%%%%%%%%%%%
%%%%%%%%%%%%%%%%%%%%%%%%%%%%%%%%%%%%%%%%%%%%%%%%%%%%%%%%%%%%%%%%%%%%%%%
%\begin{comment}
%%%%%%%%%%%%%%%%%%%%%%%%%%%%%%%%%%%%%%%%%%%%%%%%%%%%%%%%%%%%%%%%%%%%%%%
\end{document}